\documentclass[twoside]{article}
\usepackage{fleqn,espcrc2}

\newcommand{\AmS}{{\protect\the\textfont2
  A\kern-.1667em\lower.5ex\hbox{M}\kern-.125emS}}
\newcommand{\be}{\begin{eqnarray}}
\newcommand{\ee}{\end{eqnarray}}
\newcommand{\bc}{\begin{center}}
\newcommand{\ec}{\end{center}}
\newcommand{\barl}{\begin{array}{rl}}
\newcommand{\barr}{\begin{array}{rr}}
\newcommand{\ball}{\begin{array}{llllll}}
\newcommand{\ea}{\end{array}}
\newcommand{\nnb}{\nonumber}
\newcommand{\bea}{\begin{eqnarray}}
\newcommand{\eea}{\end{eqnarray}}
\newcommand{\Bdsll}{B^0_{d,s} \to l^+ l^-}

\def\be{\begin{eqnarray}}
\def\ee{\end{eqnarray}}
\def\ba{\begin{eqnarray}}
\def\ea{\end{eqnarray}}
\def\nnb{\nonumber}

\def\nnb{\nonumber}

\def\rt{\right}

\def\lt{\left}
\title{CP violation in $B_{d,s} \to l^+
l^-$ }
\author{Chao-Shang Huang (Presenting author)\address{Institute of Theoretical Physics, Academia 
        Sinica, 100080 Beijing, China},
Wei Liao\address{The Abdus Salam Internatioal Centre
for Theoretical Physics, Trieste, Italy }}

\begin{document}

\begin{abstract}
  CP violation in rare leptonic B decays are analyzed in the standard model (SM) and models
beyond SM (supersymmetric models and two Higgs doublet models). 
\vspace{1pc}
\end{abstract}

\maketitle

\section{INTRODUCTION}

The FCNC process $\Bdsll$ has been shown in recent years
to be a powerful process to shed light on new physics beyond SM~\cite{hly,bk,csb,ddn,abbt}
especially for SUSY models which may enhance the decay amplitude by
$\tan^3\beta$~\cite{hly,hpt,bk,csb}, provided that $\tan\beta$ is large (say, 
$ \geq 20$). It became more interesting recently after the Brookhaven
National Laboratory (BNL) reported the $2 \sigma$ excess of the
muon anomalous magnetic moment $a_\mu=(g-2)_\mu/2$ over its SM value:
$\Delta a_{\mu}\equiv a_\mu^{exp}-a_\mu^{SM}=2\times 10^{-9}$ at a $2\sigma$ level~\cite{BNL}. 
SUSY models with large $\tan\beta$ (say, $\geq 10$ ) are favored by this excess.
It was shown for mSUGRA in \cite{ddn}
that with a good fit to $(g-2)_{\mu}$
\footnote{The analysis in ref.\cite{ddn} was based on 
the old data, $\Delta a_{\mu}\equiv a_\mu^{exp}-a_\mu^{SM}=(43\pm 16)\times 10^{-10}$.}
  the branching ratio (Br) of $\Bdsll$
can be enhanced by 100 times and within good reach at Tevatron
Run II.

Besides Br, is there CP violation in the decays which can be observed?
In this talk I shall discuss the subject.

CP violation in the $b$-system has been established from measurements of time-dependent asymmetries in
$B \to J/\Psi K$ decays \cite{sin2betababar,sin2betabelle}. We shall show that to observe the CP asymmetry
in the B decays to a pair of muons or taus is a good way to search for new physics.
Direct CP violation is absent and no T-odd projections can be defined in the decays. However, there is
CP violation induced by $B^0-\bar{B}^0$ mixing in the process
\be
B^0\rightarrow\bar{B}^0\rightarrow f~~~~~~~vs.~~~~~~~~\bar{B}^0
\rightarrow B^0\rightarrow \bar{f}. \nnb
\ee
One can define the CP violating observable as
\bea
&&A_{CP} = \frac{D}{S},\nnb\\ && D =
\int^\infty_0 dt ~\sum_{i=1,2}\Gamma(B^0_{phys}(t) \rightarrow f_i) \nnb\\
&& -\int^\infty_0 dt ~\sum_{i=1,2}\Gamma(\bar{B}^0_{phys}(t) \rightarrow \bar{f}_i), \nnb\\
&& S = \int^\infty_0 dt ~\sum_{i=1,2}\Gamma(B^0_{phys}(t) \rightarrow f_i) \nnb\\
&& +\int^\infty_0 dt ~\sum_{i=1,2}\Gamma(\bar{B}^0_{phys}(t) \rightarrow \bar{f}_i)
\label{cp}
\eea
where $f_{1,2}=l^+_{L,R}l^-_{L,R}$ with $l_{L(R)}$ being the
helicity eigenstate of eigenvalue $-1 (+1)$, $\bar{f}_i$ is the CP conjugated state of $f_i$.

In SM, one has
\be
\frac{q}{p}= - \frac{M^*_{12}}{|M_{12}|}= - \frac{\lambda^*_t}{\lambda_t},
\label{mix}
\ee
where $\lambda_t=V_{tb} V_{td}^*$ or $V_{tb} V_{ts}^*$,
up to the correction smaller than or equal to order of $10^{-2}$, it is
\bea
A_{CP} &=& - \frac{2 Im(\xi) X_q}
{(1+ |\xi|^2)(1+X_q^2)}, ~~~q=d,s,\label{app}
\eea
where $X_q= \frac{\Delta m_q}{\Gamma}$($q=d,s$ for $B^0_d$ and $B^0_s$),
\be
\xi=
 \frac{C_{Q1}\sqrt{1-4\hat{m}_l^2}
+ (C_{Q2}+2 \hat{m}_l C_{10})}{C^*_{Q1}\sqrt{1-4\hat{m}_l^2}
- (C^*_{Q2}+2 \hat{m}_l C^*_{10})}.
\label{rate}
\ee
In eq. (\ref{rate}) $\hat{m}_l = m_l/m_{B^0}$. In SM $C_{10}$
is real, $C_{Q_2}=0$, and $C_{Q_1}$ is negligibly small. Therefore,
there is no CP violation in SM.
If  one includes the correction of order of $10^{-2}$ to $|{q \over p}|=1$, one will have  CP violation
of order of $10^{-3}$ which is unobservably small. It should be pointed out that there is no hadronic uncertainty since the common uncertain decay
constant, which is the only source of hadronic 
uncertainty for the process, cancells out in eq. (\ref{rate}). Because $C_{Q_i}$'s are
proportional to $m_l$ and $C_{10}$ is independent of $m_l$ it follows from eq. (\ref{rate}) that CP asymmetry in $B_{d,s} 
\to l^+ l^-$ is independent of the mass of the lepton, i. e., it is the same for l = electron, muon and tau, when one assumes
$\sqrt{1-4\hat{m}_l^2}\simeq 1$ which is of a good approximation for electron and muon.

In MSSM we can still use eq. (\ref{app}) in the approximation, eq. (\ref{mix}), which is a
good approximation in MSSM if one limits himself to the regions
with large $\tan\beta$ ( say, larger than 10 but smaller than 60 ), not too
light charged Higgs boson ( say, larger than 250 Gev ), and heavy sparticles,
and in the scenarios of the minimal flavor violation (MFV)\footnote{MFV means the models in
which the CKM matrix remains the unique source of flavor violation.}
 without new CP violating phases there is no correction to eq. (\ref{mix})~\cite{bcrs,ir}.
In MFV models with new CP violating phases, e.g., in the CMSSM with nonuniversal
gaugino masses which we consider in this talk, a rough estimate gives that the correction to
the SM value of $q/p$ is below $20 \%$ in the parameter space we
used in calculations~\cite{hl1}. 

>From eq. (\ref{rate}), we need to calculate the Wilson coefficients
$C_{Q_i}$ (i=1,2) and $C_{10}$ in models beyond SM in order to compute the CP asymmetry.
We shall consider the CP spontaneously broken 2HDM~\cite{hz} and the CMSSM with nonuniversal gaugino masses~\cite{nonu}. 
Compared with mSUGRA with CP violating phases
(the phase of the Higgsino mass parameter $\mu$
 and the phase of $A$), in the CMSSM with nonuniversal gaugino masses there are two more real parameters
(say, $|M_1|$ and $|M_3|$, where $M_1$ and $M_3$ are gaugino masses
corresponding to $U(1)$ and $SU(3)$ respectively) and two
more independent phases arising from complex gaugino masses, which
make the cancellations among various SUSY contributions to EDMs easier
than in mSUGRA with CP violating phases and relatively large values of the phase of $\mu$
are allowed~\cite{hl0}. The Wilson coefficients
$C_{Q_i}$ (i=1,2) and $C_{10}$ in 2HDM have been calculated respectively in ref.\cite{dhh,hz} and ref.\cite{gri}. 
In CMSSM they have also been calculated and can be found
in refs.~\cite{hly,bk,csb}\footnote{Note that
the explicit expressions of the Wilson coefficients $C_{Q_i}$'s are the same in SUSY
models with and without CP violating phases. In the SUSY models with CP violating
phases the coefficients become complex since the new CP violating phases enter into
squark and chargino mass matrices~\cite{hl1}.}. We show the leading terms of relevant Wilson coefficients in the large $\tan\beta$
case in CMSSM in the following.\\

\ba
&& C_{10} (m_W) = \nnb\\
&& \frac{m_{\ell}^2}{m_W^2} \tan^2\beta \sqrt{x_{\chi_j^-} x_{\chi_i^-}} U_{j2}^* U_{i2}
\nonumber\\
&& f_{D^0}(x_{\chi_i^-},x_{\chi_j^-},x_{\tilde u_k},x_{\tilde \nu_l}) + ...  
\ea
\ba
&& C_{Q_1}(m_W) = \nnb \\
 && - \tan^3\beta \frac{m_b m_\ell}{4 \sin^2\theta_w m_W \lambda_t} \sum_{i=1}^2
\sum_{k=1}^6 U_{i2}
T^{km}_{UL} K_{m b} \nnb \\
&& \{-\sqrt{2} V_{i1}^* (T_{UL} K)_{ks}+ V_{i2}^* \frac{ (T_{UR}
{\tilde m_u} K)_{ks}}{m_W \sin\beta} \} \nnb \\ 
&&r_{hH} \sqrt{x_{\chi_i^-}}
f_{B^0}\lt(x_{\chi_i^-},
x_{\tilde u_k}\rt) +O(\tan^2\beta),\label{c1a}\\
&& C_{Q_2}(m_W) = \nnb \\
&& \tan^3\beta \frac{m_b m_\ell}{4 \sin^2\theta_w m_W \lambda_t} \sum_{i=1}^2
\sum_{k=1}^6 U_{i2} T^{km}_{UL} K_{m b} \nonumber\\
&& \{-\sqrt{2} V_{i1}^* (T_{UL} K)_{ks}+ V_{i2}^*
\frac{(T_{UR}
{\tilde m_u} K)_{ks}}{m_W \sin\beta}\}\nonumber \\ && r_{A} \sqrt{x_{\chi_i^-}}
f_{B^0}\lt(x_{\chi_i^-},
x_{\tilde u_k}\rt) +O(\tan^2\beta),\label{c2a}
\ea
where $U$ and $V$ are matrices which diagonalize the mass matrix of charginos,
$T_{Ui}$ (i=L, R) is the matrix which diagonalizes the mass matrix of the scalar
up-type quarks and K is the CKM matrix.

The $\tan^3\beta$ enhancement of $C_{Q_i}$ (i=1,2) was first shown in ref.~\cite{hly} and confirmed later in
refs.~\cite{hpt,bk,csb}.
The chargino-chargino box diagram gives a contribution proportional to
$\tan^2\beta$ to $C_{10}$. Numerically, $C_{10}$ is enhanced in CMSSM
at most by about $10\%$ compared with SM.

In a 2HDM with CP violating phases and CMSSM with nonuniversal gaugino masses
the CP asymmetries depend on the
parameters of models
 and can be as large as $40\%$ for $B^0_d$ and $3\%$ for $B^0_s$,
  while the constraints from EDMs of electron and neutron  are satisfied~\cite{hl1}.

The correlations between
$(g-2)_{\mu}$ and CP asymmetries in $B^0_{d,s} \to l^+ l^-$ and $b \to
s \gamma$ in SUSY models with nonuniversal gaugino masses have been calculated in ref.~\cite{hl1},
imposing the constraints from the
branching ratio of $B\rightarrow X_s \gamma$ (it leads to the correlation between $C_7^{eff}$ and $C_{Q_i}$
in SUSY models) and EDMs of electron and neutron, and the results are\\
----with a good fit to the muon $g-2$ constraint,
the CP asymmetry can be as large as $25\%$ ($15\%$) for
$B^0_d \to \tau^+ \tau^-$ ($B^0_d \to \mu^+ \mu^-$) in
CMSSM with nonuniversal gaugino masses and MFV scenarios of MSSM.\\
----If tau events identified with $6\%$ tagging error, one can measure  $A_{CP}$ to a $3 \sigma$ level
at Tevatron  Run II with
Br($B_d\rightarrow \tau^+\tau^-$) enhanced by a factor of about 30  compared to that of SM.\\
----A scenario in which new physics only increases the Br a little and
the Br is still in
the uncertain region of the SM prediction. The CP asymmetry
for $B^0_d \to l^+ l^-$ in the scenario can still reach $20\%$ allowed by the muon g-2 constraint
within $2\sigma$ deviations.
So it is powerful to shed light on physics beyond SM while
the CP asymmetry of $b \to s \gamma$ in this case can only
reach $2\%$ at most which is too small to draw a definite
conclusion on new physics effects at B factories.

In summary, an observation of CP asymmetry in the decays
$B^0_q \to l^+l^- (q=d,s, l=\mu, \tau)$ would unambiguously signal the existence of new physics.


\begin{thebibliography}{9}
\bibitem{hly}C.-S.~Huang and Q.-S.~Yan, Phys. Lett. {\bf B442} (1998) 209; C.-S.~Huang, W.~Liao and Q.-S.~Yan,
Phys.\ Rev.\ {\bf D59} (1999) 011701. 
\bibitem{bk}K.S. Babu and C. Kolda, Phys. Rev. Lett. {\bf 84} (2000) 228.
\bibitem{csb}C.-S. Huang et al., Phys. Rev. {\bf D63} (2001) 114021; ibid. {\bf 64} (2001) (2001) 059902(E); 
P.H. Chankowski, L. Slawianowska, P.R. {\bf D63} (2001) 054012; C. Bobeth et al., Phys. Rev. {\bf D64}
(2001) 074014; hep-ph/0204225; G. Isidori, A. Retico, JHEP {\bf 11} (2001) 001.
\bibitem{ddn} A. Dedes, H. K. Dreiner and U. Nierste, Phys. Rev. Lett. {\bf 87} (2001) 251804.
\bibitem{abbt}R. Arnowitt et al., hep-ph/0203069; S. Baek, P. Ko and W. Y. Song, hep-ph/0205259; H. Dreiner, 
U. Nierste and P. Richardson, hep-ph/0207026; J. K. Mizukoshi, X. Tata and Y. Wang, hep-ph/0208078. For a recent
review, see, e.g., Chao-Shang Huang, hep-ph/0210314.
\bibitem{hpt}C. Hamzaoui, M. Pospelov and M. Toharia, Phys. Rev. {\bf D59} (1999) 095005.
\bibitem{BNL} G.W. Bennet {\it et al.}, 
hep-ex/0208001.
\bibitem{sin2betababar}
B.~Aubert {\it et al.}  [BABAR Collaboration],
Phys.\ Rev.\ Lett.\  {\bf 87}, 091801 (2001).
\bibitem{sin2betabelle}
K.~Abe {\it et al.}  [Belle Collaboration],
Phys.\ Rev.\ Lett.\  {\bf 87}, 091802 (2001).
\bibitem{bcrs}A.J. Buras et al., hep-ph/0107048.
\bibitem{ir}G. Isidori and A. Retico, hep-ph/0110121.
\bibitem{hz}C.-S. Huang and S.-H. Zhu, Phys. Rev. {\bf D61} (2000) 015011, Erratum-ibid. {\bf D61} (2000) 119903.
Chao-Shang Huang, Liao Wei, Qi-Shu Yan, Shou-Hua Zhu,
 Euro. Phys. J. {\bf C25} (2002) 103.
\bibitem{nonu}A. Brignole, L. Ib\'a\~nez, C. Mu \~noz, Nucl. Phys. {\bf B422}
(1994) 125, ibid. {\bf B436} (1995) 747(E);
M.Brhlik, G.J.Good, and G.L.Kane, Phys. Rev. {\bf D59} 11504(1999); N. Chamoun,
C.-S. Huang, C. Liu and X.-H. Wu, Nucl. Phys. {\bf B624} (2002) 81.
\bibitem{dhh}Y.-B. Dai, C.-S. Huang, and H.-W. Huang, Phys. Lett. {\bf B390} (1997) 257;
 ibid. {\bf B513} (2001) 429(E); H.E. Logan and U. Nierste,  Nucl. Phys. {\bf B586} (2000) 39.
\bibitem{gri}B. Grinstein, M.J. Savage, M.B. Wise, Nucl. Phys. {\bf B319} (1989) 271.
\bibitem{hl1}C.-S. Huang, W. Liao, Phys. Lett. {\bf B525} (2002) 107;
 Phys. Lett. {\bf B538} (2002) 301.
\bibitem{hl0}T. Ibrahim, P. Nath, Phys. Lett {\bf B418} (1998) 98;
 C-S. Huang, W. Liao, Phys. Rev. {\bf D62} (2000) 016008.
\end{thebibliography}
\end{document}